# Female citation impact superiority 1996-2018 in six out of seven English-speaking nations[1]

Mike Thelwall, Statistical Cybermetrics Research Group, University of Wolverhampton, UK.

Efforts to combat continuing gender inequalities in academia need to be informed by evidence about where differences occur. Citations are relevant as potential evidence in appointment and promotion decisions, but it is unclear whether there have been historical gender differences in average citation impact that might explain the current shortfall of senior female academics. This study investigates the evolution of gender differences in citation impact 1996-2018 for six million articles from seven large English-speaking nations: Australia, Canada, Ireland, Jamaica, New Zealand, UK, and the USA. The results show that a small female citation advantage has been the norm over time for all these countries except the USA, where there has been no practical difference. The female citation advantage is largest, and statistically significant in most years, for Australia and the UK. This suggests that any academic bias against citing female authored research cannot explain current employment inequalities. Nevertheless, comparisons using recent citation data, or avoiding it altogether, during appointments or promotion may disadvantage females in some countries by underestimating the likely impact of their work, especially in the long term.
**Keywords**: Citation impact; gender differences; research evaluation; academic careers.

## Introduction

There is a long history of overt and covert discrimination against females in universities and society. Despite formal gender equality in employment being half a century old in many countries (e.g., the US Civil Rights Act of 1964; the UK Sex Discrimination Act 1975), females have not yet achieved parity in academia in most countries (Larivière, Ni, Gingras, Cronin, & Sugimoto, 2013). Current imbalances include fewer females overall (Shannon, Jansen, Williams, Cáceres, Motta, et al., 2019) and in most fields (Thelwall, Bailey, Makita, Sud, & Madalli, 2019; Thelwall, Bailey, Tobin, & Bradshaw, 2019) as well as even lower proportions of females in promoted posts (e.g., Bosquet, Combes, & García-Peñalosa, 2018). Gender imbalances seem to be decreasing (e.g., Winchester & Browning, 2015) but are an ongoing concern and may not disappear in our lifetimes (Holman, Stuart-Fox, & Hauser, 2018).

     National initiatives that publicly attempt to redress the balance (Tzanakou & Pearce, 2019; Van Miegroet, Glass, Callister, & Sullivan, 2019) are unlikely to succeed unless the reasons for the imbalances are understood. The causes of low proportions of senior female academics have been debated, with bias from male colleagues (De Paola & Scoppa, 2015; Ooms, Werker, & Hopp, 2019) or systemic biases (Nielsen, 2015; Rivera, 2017) being possible explanations for the leaky pipeline of female academic careers in some fields (Clark Blickenstaff, 2005). Chilly climates for females may hamper entry into disciplines (Britton, 2017) and there have been efforts to combat this (Stockard, Greene, Richmond, & Lewis, 2018). Bias against female scientists by avoiding citing their work has previously been hypothesised to contribute to a lack of females in senior positions (Larivière, Ni, Gingras, Cronin, & Sugimoto, 2013; Paul-Hus, Bouvier, Ni, Sugimoto, Pislyakov, & Larivière, 2015), but

---

it has recently been found that female-authored research in the USA and UK is more cited (Elsevier, 2017; Thelwall, 2018a). This occurs despite a higher self-citation rate for male authors (Deschacht & Maes, 2017; King, Bergstrom, Correll, Jacquet, & West, 2017). Given that current gender imbalances in seniority are the result of career-long factors, it is important to know whether the 2014 female citation advantage is a recent trend or a long-term factor. One longitudinal study has investigated six fields in the USA 1996-2017, finding no evidence for the cause-and-effect hypothesis that higher average citation rates for one gender led to an increase in the proportion of researchers from that gender (Thelwall, 2018c). More systematic evidence is needed, however.

Many articles have investigated gender differences in citation rates for individual fields, finding that male-authored articles are cited more (Dion, Sumner, & Mitchell, 2018; Gewin, 2017; Maliniak, Powers, & Walter, 2013; Schisterman, Swanson, Lu, & Mumford, 2017), the same (Barrios, Villarroya, & Borrego, 2013; Borsuk, Budden, Leimu, Aarssen, & Lortie, 2009; Copenheaver, Goldbeck, & Cherubini, 2010; Lynn, Noonan, Sauder, & Andersson, 2019) or less (Cotropia & Petherbridge, 2017) than female-authored articles, depending on the field. At the level of scholars, males tend to have more career citations (e.g., Aksnes, Rorstad, Piro, & Sivertsen, 2011; Beaudry & Larivière, 2016; Geraci, Balsis, & Busch, 2015; Porter, 2018), at least partly because they are less likely to work part-time or take career gaps for carer responsibilities (Ceci, Williams, & Barnett, 2009; Ceci & Williams, 2011; Jappelli, Nappi, & Torrini, 2017; see also: Evans, 2016).

A global analysis of the gender of authors of scientific articles 2008-2012 in the Web of Science found almost all countries to have mostly male authored articles, presumably due to gender differences in employment more than differences in career outputs. Countries with the most research output also had a citation advantage for male authors (Larivière, Ni, Gingras, Cronin, & Sugimoto, 2013). A report by Elsevier using Scopus 1996-2015 confirmed the dominance of male-authored papers in 11 out of 12 countries or regions (exception: Japan) and a citation advantage for male first-authored papers 2011-15 for the EU, Canada, Australia but a citation advantage for female first-authored papers 2011-15 for the UK and the USA. Gendered citation differences 1996-2010 were compared to 2011-15, with only Canada reversing (from female advantage to male advantage) (Elsevier, 2017). The change evidence was not fine-grained enough to reveal any general patterns, however, except that gender difference directions (favouring male or female) were relatively stable, although their magnitude was not. Two studies using a more robust method of averaging citations the mean normalised log citation score (MNLCS, discussed in the methods section) have found evidence of slightly higher impact for female first-authored research for the USA, UK and Spain, but a male advantage for India and Turkey (Thelwall, 2018a). This article used Scopus journal articles from 2014 but the difference in outcomes between this and one of the previous papers (Larivière, Ni, Gingras, Cronin, & Sugimoto, 2013) was due to the use of the MNLCS field normalisation indicator in the later paper (Thelwall, 2018a) that is more suitable for highly skewed citation data. The underlying reason for this making a difference is that males seem to write a greater fraction of the few extremely highly cited articles (Baltussen & Kindler, 2004; Graham, Pratt, Lee, & Cullen, 2019; Schisterman, Swanson, Lu, & Mumford, 2017; Wong, Tan, & Sabanayagam, 2019) that dominate the average (as previously noted by: Zigerell, 2015). These highly cited papers are more likely to make a methods contribution (Boyack, van Eck, Colavizza, & Waltman, 2018; Small, 2018), such as introducing a widely used computer program. Thus, they may not always make the most important contributions to progress in a field, even if they are very useful and sometimes used beyond academia.

There may be a gender bias against citing females even if they are cited more, if female authored research is much more impactful than any citation difference suggests. This is a distinct possibility because females in the USA, and probably other countries, are more likely to choose careers with a positive societal impact (Diekman, Steinberg, Brown, Belanger, & Clark, 2017) and so their research might naturally be more useful. This would be difficult to assess empirically, but one study has suggested that female first-authored research generates more educational impact per citation compared to male first-authored research in Spain, Turkey, UK and USA, but conversely for India (Thelwall, 2018b). This evidence came from users of the social reference sharing site Mendeley that were registered as students.

Many studies have claimed that male researchers within a country or field are more productive, in the sense of writing more journal articles (Aksnes, Rorstad, Piro, & Sivertsen, 2011; Ceci, Ginther, Kahn, & Williams, 2014; Nielsen, 2016; Raj, Carr, Kaplan, Terrin, Breeze, & Freund, 2016; Rørstad & Aksnes, 2015). If true, this might explain the dearth of female appointments and promotions in some fields. These apparent differences can disappear or greatly reduce when studies take into account the greater degree of part-time working and additional teaching commitments of female staff (Ceci, Ginther, Kahn, & Williams, 2014; Cameron, White, & Gray, 2016; van den Besselaar, & Sandström, 2017; Xie & Shauman, 1998). Given that no previous study seems to have undertaken a comprehensive large-scale study of gender and productivity in academia and there are multiple sources of evidence that female productivity can be accidentally underestimated by failing to take into account the amount of female academic staff time that is available for research, it is not possible to conclude that there is a productivity gap in terms of the number of papers written in the time available. Nevertheless, male researchers may tend produce more articles over their career since they are less likely to take career breaks for caring responsibilities or work part-time or shorter hours for a period (McMunn, Lacey, Worts, McDonough, Stafford, et al., 2015). The extent to which this occurs varies internationally (Sieverding, Eib, Neubauer, & Stahl, 2018). Promotion, funding and tenure committees may sometimes fail to fully take into account career circumstances, disadvantaging any female applicants with shorter CVs due to career breaks or part-time working.

This article assesses the history of gender differences in average citation rates for seven large mainly English-speaking countries with overlapping cultures: Australia, Canada, Ireland, Jamaica, New Zealand, UK, USA. This selection was chosen to allow comparisons between the countries, without problems caused by extensive non-English language publishing. Large countries were chosen to give clearer patterns and a chance of statistically significant results. This paper follows a (partly) longitudinal comparison report (Elsevier, 2017) with more recent data, a different set of countries, a more robust field weighted citation indicator, finer-grained results, statistical evidence, and information about the whole distribution of citations (relevant to the hypothesised influence of highly cited articles).

- RQ1: How have gender imbalances in publishing evolved over time in each country?
- RQ2: How have gender imbalances in citation impact evolved over time in each country?
- RQ3: What is the distribution of citation impact, by gender, in each country?

# Methods

Scopus-indexed journal articles 1996-2018 were used to address the research questions. Reviews and other non-article outputs were excluded to focus on standard primary research in most fields. Although books and conference papers are the main outputs in some disciplines (Engels, Ossenblok, & Spruyt, 2012; Goodrum, McCain, Lawrence, & Giles, 2001; Norris & Oppenheim, 2003), these are less comprehensively indexed than journal articles and so were not included. Moreover, citations probably have little value for evaluating the impact of books. Scopus was used rather than the Web of Science because it has wider coverage of academic research (Mongeon & Paul-Hus, 2016) and finer-grained set of 334 subject categories (Elsevier, 2019). The period 1996-2018 was covered because Scopus expanded in 1996 and 2018 is the most recent complete year. The citation counts for 1996-2017 were downloaded in November and December 2018 and the citation counts for 2018 at the end of January 2019.

The set of articles for each country consisted of documents of the Scopus Journal Article type, published 1996-2018, and with the first author affiliation from that country, as recorded in Scopus.

## Gender detection

The first author of each paper was assumed to be the main contributor in all cases. Whilst this is true in general in all broad fields (Larivière, Desrochers, Macaluso, Mongeon, Paul-Hus, & Sugimoto, 2016; see also: Yang, Wolfram, & Wang, 2017), alphabetical authorship occurs to some extent in some narrow fields and may even be the norm in some (Henriksen, 2019; Waltman, 2012). In cases of alphabetical authorship with *n* authors, the main author has a 1/*n* chance of being listed first by accident and (depending on the gender composition of the field and the gender of the first author) may have the same gender as the first author if not. This is more likely to occur if there is a degree of gender homophily in authorship (e.g., Fox, Ritchey, & Paine, 2018). Two or more authors are sometimes also credited with having contributed equally to a study. This seems to be most prevalent in high impact medical journals, such as New England Journal of Medicine (NEJM), where it increased from under 1% in 2000 to 8.6% in 2009 (Akhabue & Lautenbach, 2010). In NEJM, 60% of the equal credit assignments apply to the first two authors. Equal authorship has a smaller effect than alphabetical authorship because it seems to be rarer (there are apparently no large-scale studies), usually results in at least 50% of the gender assignment being correct (two authors with equal credit), and the remaining 50% has the same chance as being correct as for alphabetical authorship. If it is most prevalent for high impact medical journals, then it might have most influence on highly cited medical articles.

The gender of the first author of a paper was inferred from their first name, when present, and when there was evidence that their first name referred to males 90% or more of the time or females 90% of the time in their country of origin. The gendered name list was taken from the 1990 US census and then augmented by calls to GenderAPI.com for the gender of the first name of the first authors in the corpus, using their country affiliations. GenderAPI.com estimates first name genders from social media profiles associated with a country. It reports the proportion of times the name is associated with males or females, alongside the number of examples checked. Evidence of gender was used if a name was 100% one gender with at least 10 examples, increasing the evidence requirement as the percentage decreased, eventually falling to 90% one gender needing 500 examples. This is a more relaxed requirement than a previous study (Thelwall, Bailey, Tobin, & Bradshaw, 2019), increasing the number of names used. Many of the new names were from ethnic minorities within each country (e.g., Greek names in the UK, including Konstantinos), and Jamaican names that are rare in academia elsewhere (e.g., Linford). Nevertheless, many rare and unisex (or differently gendered amongst communities represented in the countries examined, including Nicola) were not included. This method produced almost ten million gendered journal articles, with two thirds of articles being assigned a first author gender (Table 1). The exceptions used initials, a relatively unisex first name, a rare first name or a first name from a minority culture.

Table 1. The number of gendered Scopus journal articles 1996-2018.

| Country | Articles | Gendered (%) |
| --- | --- | --- |
| Australia | 640026 | 411787 (64%) |
| Canada | 813623 | 528753 (65%) |
| Ireland | 82838 | 51442 (62%) |
| Jamaica | 3273 | 1624 (50%) |
| New Zealand | 110714 | 70704 (64%) |
| UK | 1494347 | 918493 (61%) |
| USA | 6051195 | 4058727 (67%) |
| **Total** | **9196016** | **6041530 (66%)** |

## Citation impact

Formulae derived from the numbers of citations to an academic article are often referred to as citation impact indicators on the basis that counts of citations are an approximate indicator of the extent to which an article has been found useful in future research. The average citation impact of research by each gender was calculated using the mean normalised log citation score (MNLCS) to normalise for the field and year of publication (Thelwall, 2017). It is important to normalise by field and year because the average citation counts of articles differ greatly between fields and years. In addition, the proportion of females differs between fields (greatly) and years, and so comparisons of raw citation counts would give misleading gender differences. The log variant rather than the mean normalised citation score (MNCS) (Waltman, van Eck, van Leeuwen, Visser, & van Raan, 2011ab) was used because sets of citation counts are highly skewed and the arithmetic mean is more precise after the log transformation (Thelwall & Fairclough, 2017). This also allows confidence intervals to be calculated to estimate the precision of the results.

For the MNLCS calculations, the citation count $C$ for each article in the dataset was first replaced by the log-transformed value $ln(C + 1)$ to greatly reduce skewing. The plus one in the calculation allows uncited articles to be retained, since the log of zero is undefined. Next, for each Scopus narrow field (up to 334) and year, these log-transformed citation counts were averaged with the arithmetic mean, giving $A_{f,y}$. Normalising by narrow fields is preferable to normalising by broad field because broad fields can mix high and low citation specialisms, giving an unfair advantage to the high citation specialism researchers. For example, within Chemical Engineering, the Colloid and Surface Chemistry specialism can attract four times as many citations per paper as Chemical Health and Safety. The denominator is based on articles from all countries (not just the seven analysed here), giving adequate numbers in each case. The log-transformed citation count for each article was then divided by this average for the field and year in which the article was published, giving the final normalised citation score for each article of $ln(C + 1)/A_{f,y}$. The MNLCS average citation score for any set of articles is then the arithmetic mean of the corresponding $ln(C + 1)/A_{f,y}$ values.

Except for the most recent three years, the skewness and excess kurtosis of the normalised citation counts were almost always below 3. Thus, the confidence intervals, calculated from the normalised citation counts with the standard normal distribution formula ($\bar{x} \pm 1.96s/\sqrt{n}$, where *n* is the sample size, *s* is the sample standard deviation and $\bar{x}$ is the sample mean; replacing 1.96 with the t distribution value for small sample sizes), are reliable before 2015 despite being derived from discrete skewed data before the transformations.

### *Citation distribution*

The distribution of levels of citation impact by gender was investigated by ranking all the articles by their normalised log citation score (NLCS) and then calculating the percentage of female first-authored papers with a NLCS value at least as large as every NLCS value in the set. This is similar to calculating the proportion of females in the top 1%, top 2% etc. but is more fine-grained. The NLCS for an article is its logged citation count $ln(1 + c)$ divided by the average (arithmetic mean) logged citation count $\overline{ln(1 + c)}$ for all journal articles from the same Scopus narrow field and year (MNLCS is the average of all NLCS).

## Results

The proportion of female first-authored journal articles increased 1996-2018 in all countries (Figures 1-7). This increase is steady except for Jamaica, probably due to the smaller Jamaican sample sizes for each year. The exact proportions are not reliable indicators of gendered shares of publications because the accuracy of the gender identification heuristic may vary between genders and between countries. Nevertheless, Jamaica seems to have achieved gender parity in journal article publishing, Australia seems to be approaching gender parity, and the UK seems to be slowing down well short of parity (40%). For comparison, the least gender unequal countries according to the UNDP 2017 gender inequality index (hdr.undp.org/en/composite/GII) are Canada (0.092), Australia (0.109), Ireland (0.109), UK (0.116), New Zealand (0.136), USA (0.189) and Jamaica (0.412), which does not explain the low proportion for the UK or the high proportion for Jamaica. World Economic Forum (WEF, 2018) gender inequality estimates (best to worst: New Zealand, Ireland, UK, Canada, Australia, Jamaica, USA) also do not align with the proportion results here. Thus, the gender proportion differences seem to be specific to academia within these countries rather than being reflections of national gender inequalities. The proportions may partly reflect the extent to which a nation's gendered professions, and particularly nursing, are taught in higher education institutions by people that are expected to publish research, and with that research appearing in periodicals indexed by Scopus.

With the partial exception of the USA, there is a general trend for female first-authored research to be more cited than male first-authored research in all seven countries and all years 1996-2014 (Figures 1-7). When the error bars do not overlap, the difference can be assumed to be statistically significant with p<0.05. Whilst statistical significance is also consistent with a small overlap in confidence intervals, this is not an important distinction because of several methods limitations (see Discussion). The female citation advantage most of the time is statistically significant for Australia and the UK. For Canada,

Ireland, Jamaica and New Zealand there is a female citation advantage in enough years to be confident that it is a trend, despite not being statistically significant in most or all years. For the USA any gender citation advantage is very slight, especially compared to variations over time.

There is a trend for a relative male citation advantage increase (i.e., the male MNLCS line height to increase relative to the female MNLCS line height) for 2014-18 in four countries: Australia; Canada; UK and USA. These could be due to the higher level of male self-citation since early citations are rare (and hence more influential in the MNLCS calculation) and are more likely to be self-citations because an author knows their own work before anyone else and may work on a series of related papers.

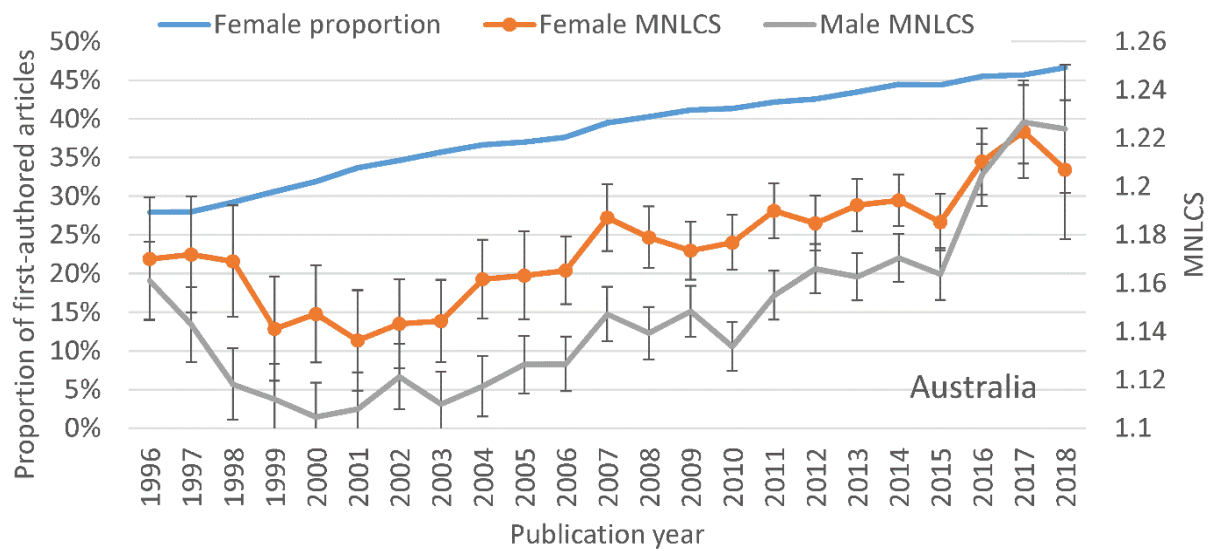

Figure 1. The percentage of **Australian** Scopus-indexed journal articles with a female first author (out of all gendered articles) and the average field normalised citation impact of gendered first-author research. Error bars reflect 95% confidence intervals and are unreliable after 2014.

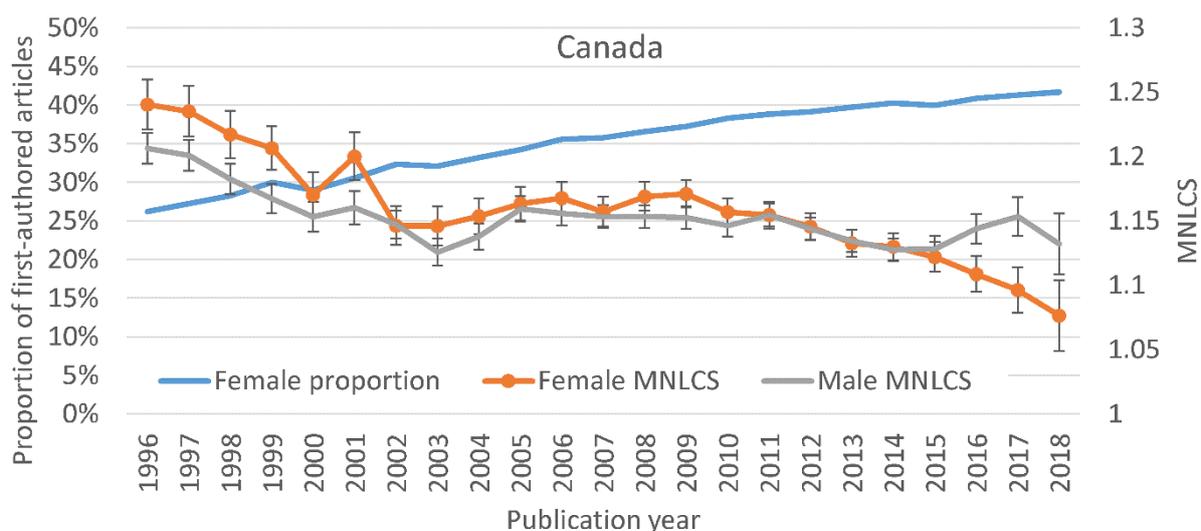

Figure 2. The percentage of **Canadian** Scopus-indexed journal articles with a female first author (out of all gendered articles) and the average field normalised citation impact of gendered first-author research. Error bars reflect 95% confidence intervals and are unreliable after 2014.

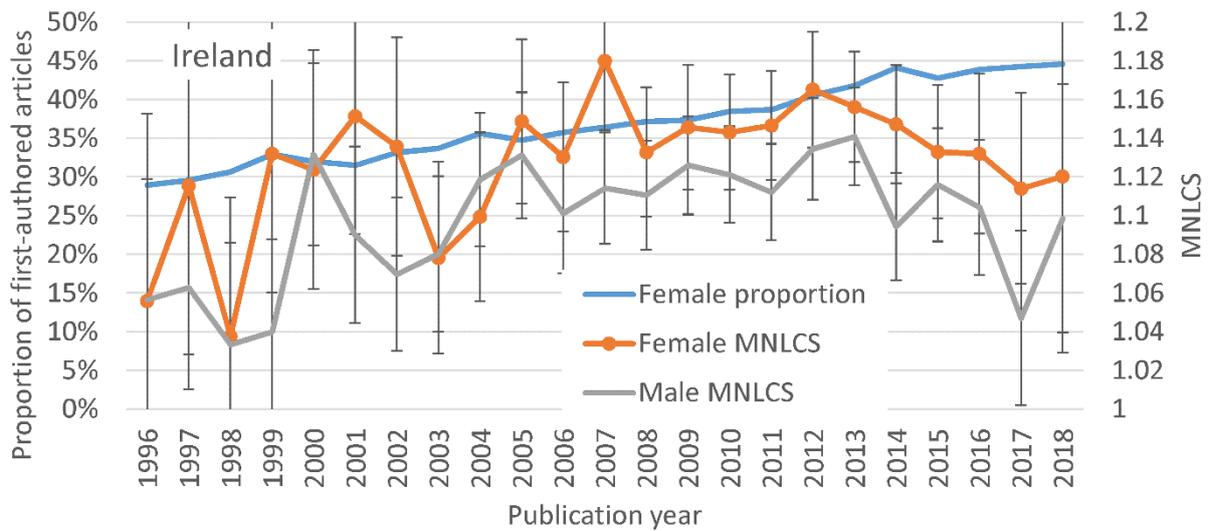

Figure 3. The percentage of **Irish** Scopus-indexed journal articles with a female first author (out of all gendered articles) and the average field normalised citation impact of gendered first-author research. Error bars reflect 95% confidence intervals and are unreliable after 2014.

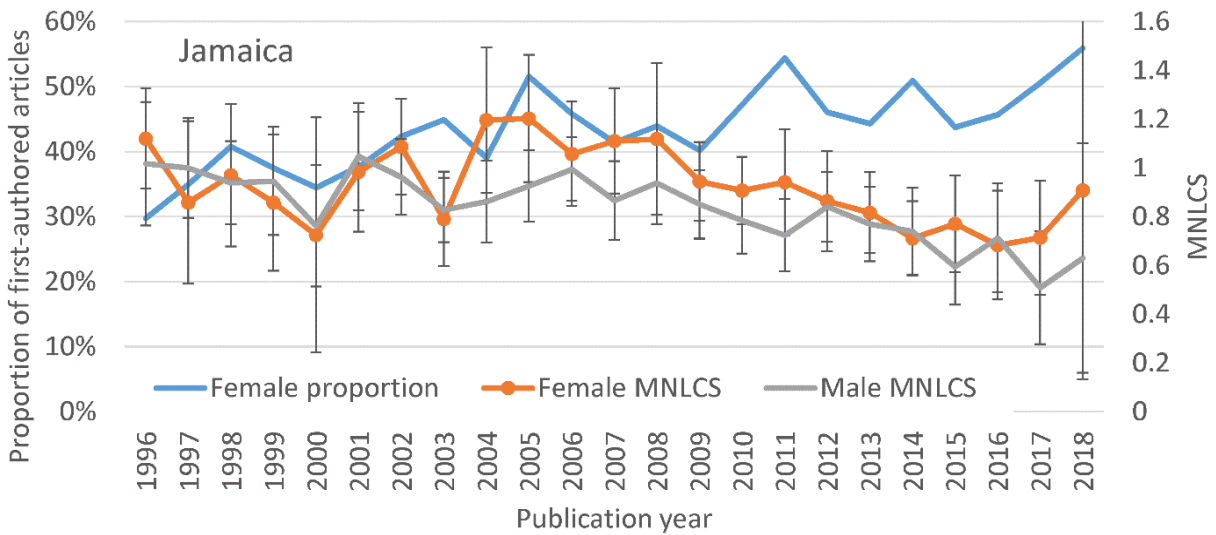

Figure 4. The percentage of **Jamaican** Scopus-indexed journal articles with a female first author (out of all gendered articles) and the average field normalised citation impact of gendered first-author research. Error bars reflect 95% confidence intervals and are unreliable after 2014.

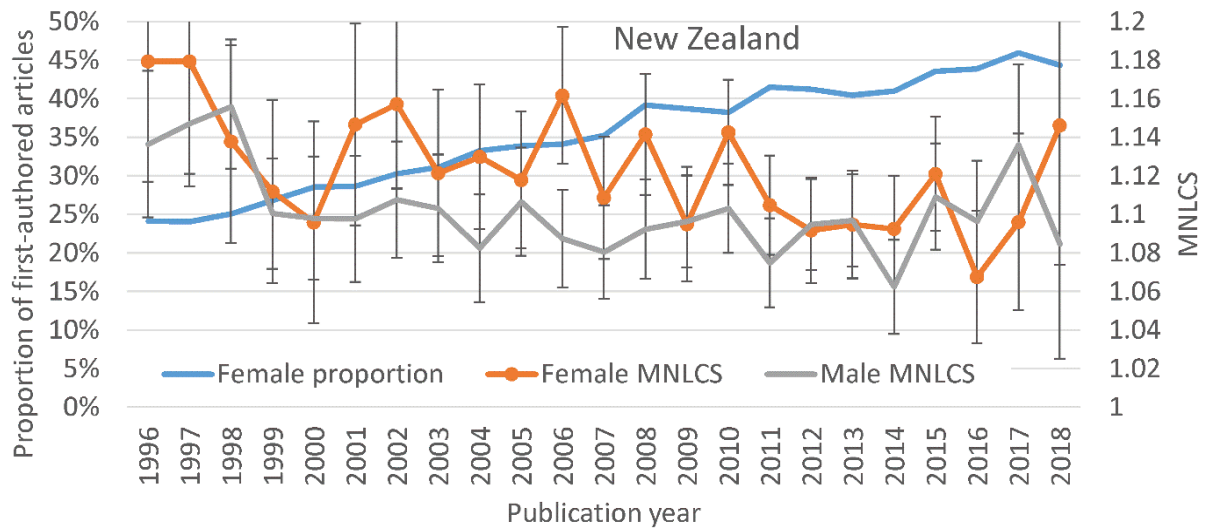

Figure 5. The percentage of **New Zealand** Scopus-indexed journal articles with a female first author (out of all gendered articles) and the average field normalised citation impact of gendered first-author research. Error bars reflect 95% confidence intervals and are unreliable after 2014.

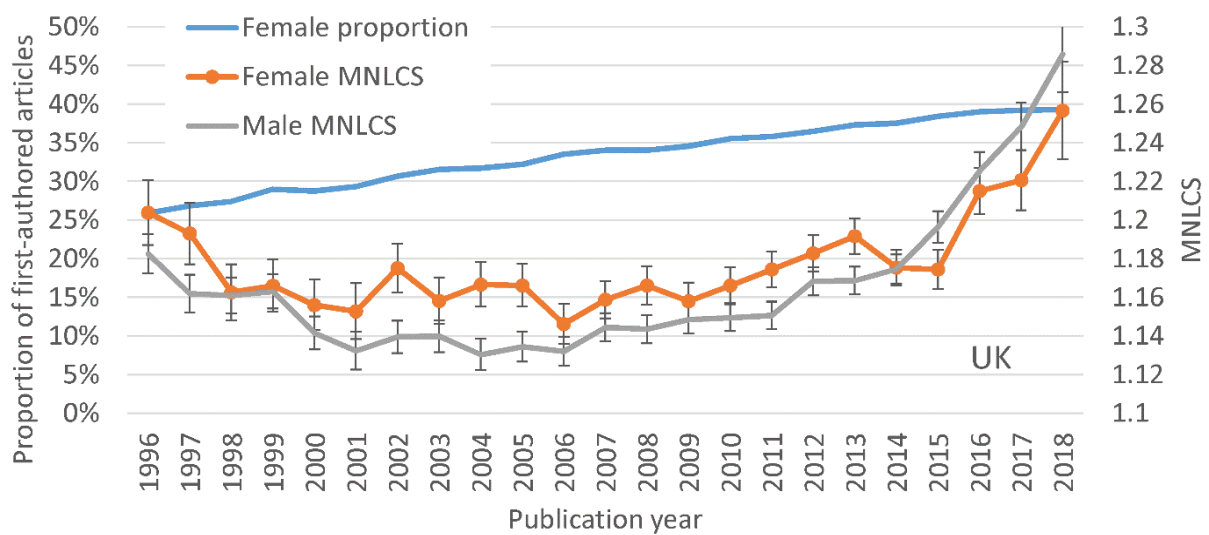

Figure 6. The percentage of **UK** Scopus-indexed journal articles with a female first author (out of all gendered articles) and the average field normalised citation impact of gendered first-author research. Error bars reflect 95% confidence intervals and are unreliable after 2014.

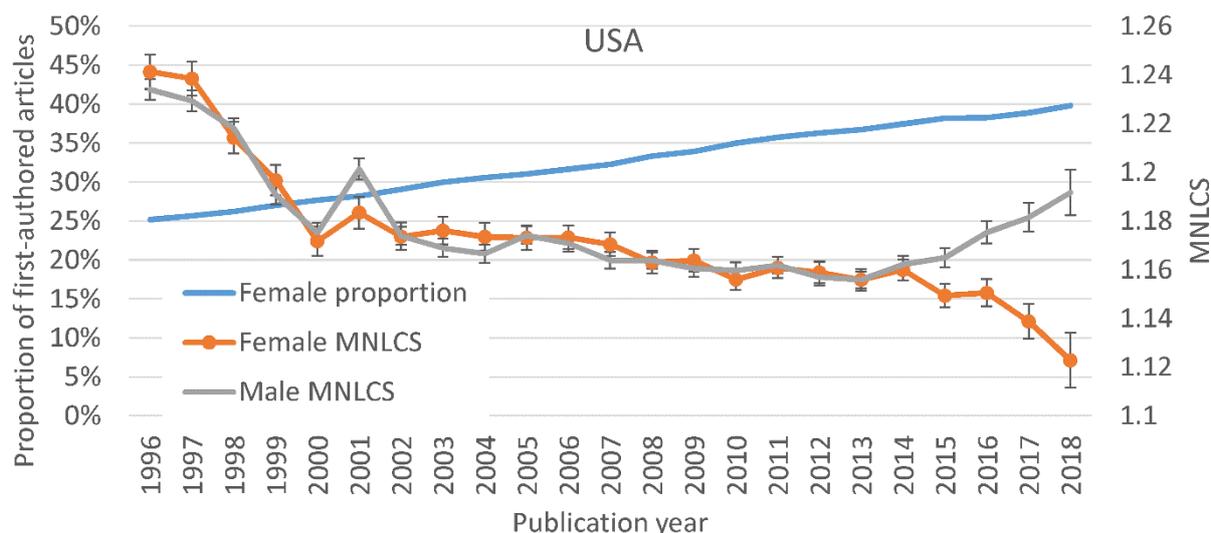

Figure 7. The percentage of **US** Scopus-indexed journal articles with a female first author (out of all gendered articles) and the average field normalised citation impact of gendered first-author research. Error bars reflect 95% confidence intervals and are unreliable after 2014.

The overall average for males and females summarises the trends evident in the graphs (Table 2).

Table 2. The average field normalised citation impact (MNLCS) of gendered first-author research for Scopus journal articles 1996-2018 and 1996-2015. The period 1996-2015 is more reliable, excluding the anomalous recent three years with a too small citation window.

| Country | 1996-2018 | | | 1996-2015 | | |
|---|---|---|---|---|---|---|
| | **Female** | **Male** | **F – M** | **Female** | **Male** | **F – M** |
| Australia | 1.1855 | 1.1575 | 0.0280 | 1.1767 | 1.1428 | 0.0339 |
| Canada | 1.1453 | 1.1501 | -0.0048 | 1.1586 | 1.1515 | 0.0071 |
| Ireland | 1.1360 | 1.1044 | 0.0317 | 1.1404 | 1.1095 | 0.0309 |
| Jamaica | 0.9160 | 0.8275 | 0.0885 | 0.9420 | 0.8559 | 0.0862 |
| New Zealand | 1.1150 | 1.0996 | 0.0154 | 1.1184 | 1.0983 | 0.0201 |
| UK | 1.1828 | 1.1713 | 0.0116 | 1.1706 | 1.1549 | 0.0157 |
| USA | 1.1632 | 1.1712 | -0.0079 | 1.1698 | 1.1703 | -0.0005 |

## Citation impact distribution by gender

The top cited percentiles (Figures 8-14) give some context to the gender difference results. The right-hand dot on each graph (above the x axis value of 100%) indicates the overall proportion of gendered articles with a female first author. The size of the gap to the left of the dot in each case is the percentage of uncited articles from the country. For example, 13% of Australian gendered articles were uncited. Points on the graph above the right-hand dot indicate that a higher proportion of females was in the top cited percentile for the country. For example, the y axis value above 20% on the x axis is 42%, indicating that 42% of the most highly cited 20% of Australian papers had a female first author. Since the overall figure for Australia is 41%, this indicates that an additional 1% of the top cited 20% of papers had a female author than overall for Australia.

In four cases (Australia, Canada, Ireland, USA), a few extremely highly cited articles are more likely to be male-authored than overall, as indicated by a thick line of dots on the y axis just above the x axis. For the largest case, the USA, this bar contains 205 articles, representing the top cited 0.005%, and so can safely be ignored.

In four cases (Australia, Ireland, Jamaica, New Zealand), the line has a consistently negative slope of decreasing magnitude and an overall similar shape. In these countries, the higher the citation impact of a paper, the more likely it is to have a female first author. This trend is strongest for the top 10% of articles.

In three cases (Canada, UK, USA), the percentile line has a more complex shape. The most highly cited few percent (e.g., 0.005%-3.5% for the USA) has more female first-authored articles than overall for the country. A large section of the next top cited articles (up to 60% for the USA) then has more male first-authored articles than overall for the country. Finally, there is an increase from the penultimate point on each graph to the final point above 100% on the x axis, indicating that more female first-authored articles are uncited than male first-authored articles. Thus, females in this group are characterised by their ability to write very highly cited articles (e.g., top 0.005%-3.5%) and an increased chance of being uncited whereas males are characterised by an increased chance to write highly cited articles (e.g., top 20%). The larger overall share of uncited articles for females is mainly due to the most recent years (and 2018 in particular), suggesting that females are more likely to publish in slower citing fields, although it may reflect their lower self-citation rates (which would be most relevant to the newest articles).

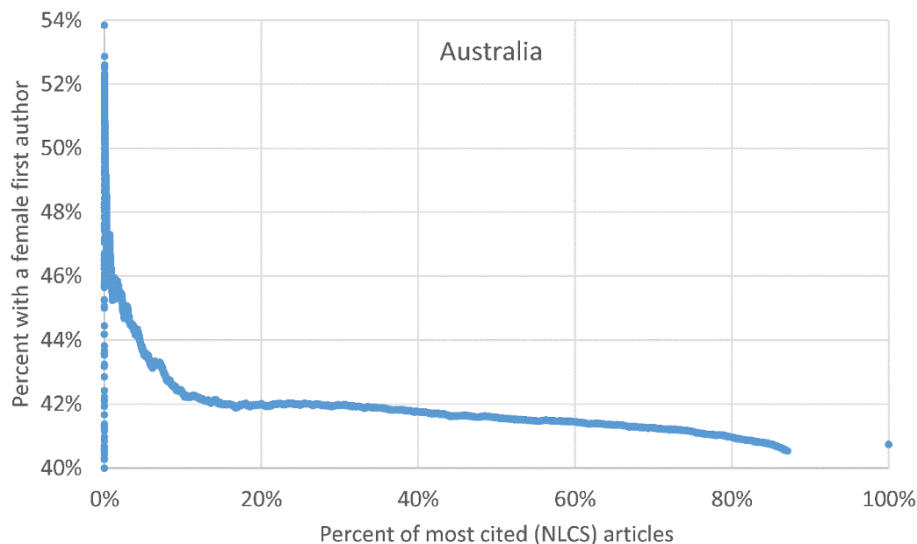

Figure 8. The cumulative percentage of female first authored **Australian** Scopus-indexed journal articles 1996-2018 in each top cited percentile (using NLCS values). For each dot on the graph, the x axis value is a cumulative percent of top-cited values and the y axis value is the female authored percentage. For example, point with x coordinate 10% has y coordinate 42.5%. This indicates that 42.5% of the 10% most cited articles have a female first-author, which is above average for Australia. The isolated dot on the right hand side with X coordinate 100% and y coordinate 40.7% corresponds to the female percentage for all Australian journal articles, which is 40.7% overall. The gap between this point and the rest of the graph corresponds to the percentage of uncited articles.

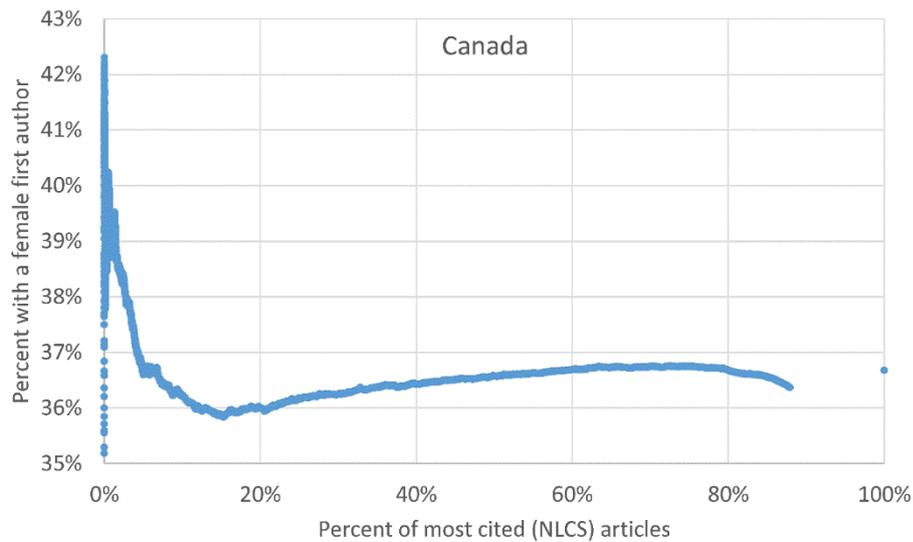
Figure 9. The cumulative percentage of female first authored **Canadian** Scopus-indexed journal articles 1996-2018 in each top cited percentile (using NLCS values).

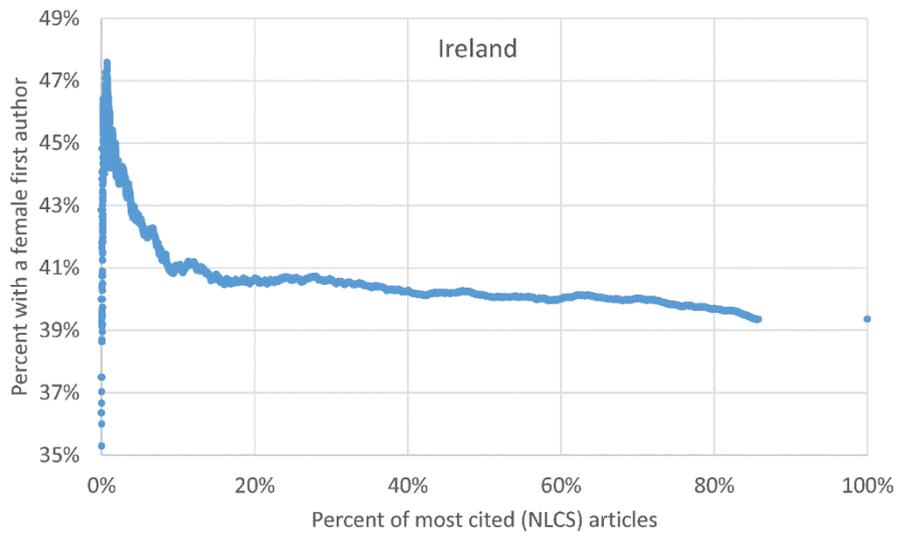
Figure 10. The cumulative percentage of female first authored **Irish** Scopus-indexed journal articles 1996-2018 in each top cited percentile (using NLCS values).

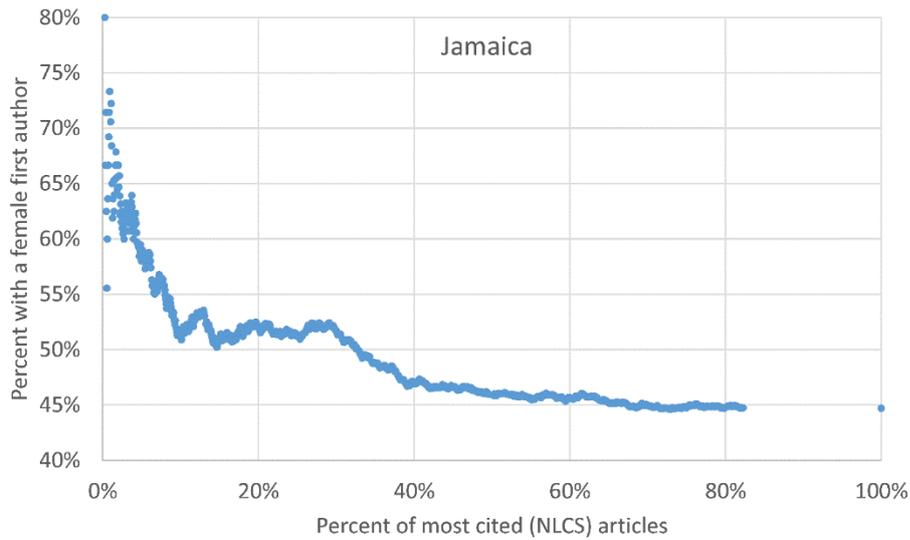

Figure 11. The cumulative percentage of female first authored **Jamaican** Scopus-indexed journal articles 1996-2018 in each top cited percentile (using NLCS values).

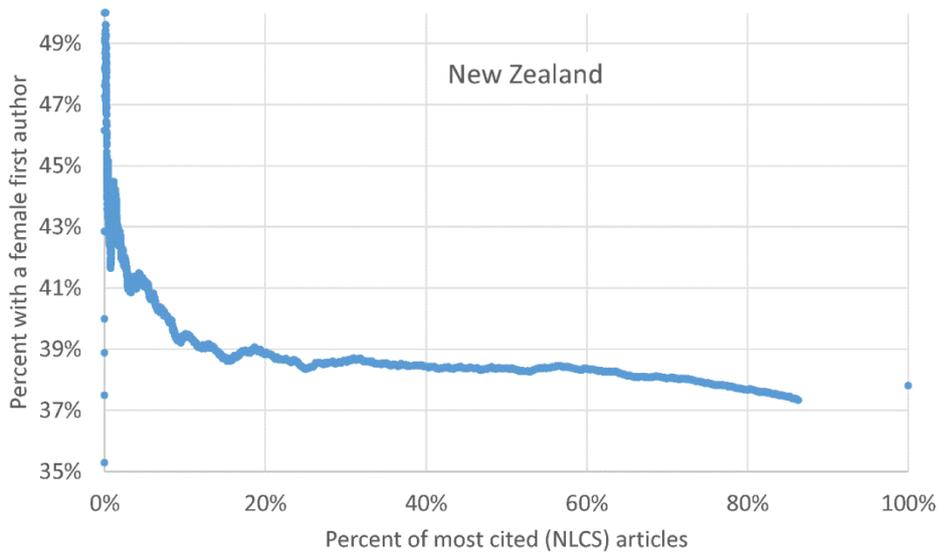

Figure 12. The cumulative percentage of female first authored **New Zealand** Scopus-indexed journal articles 1996-2018 in each top cited percentile (using NLCS values).

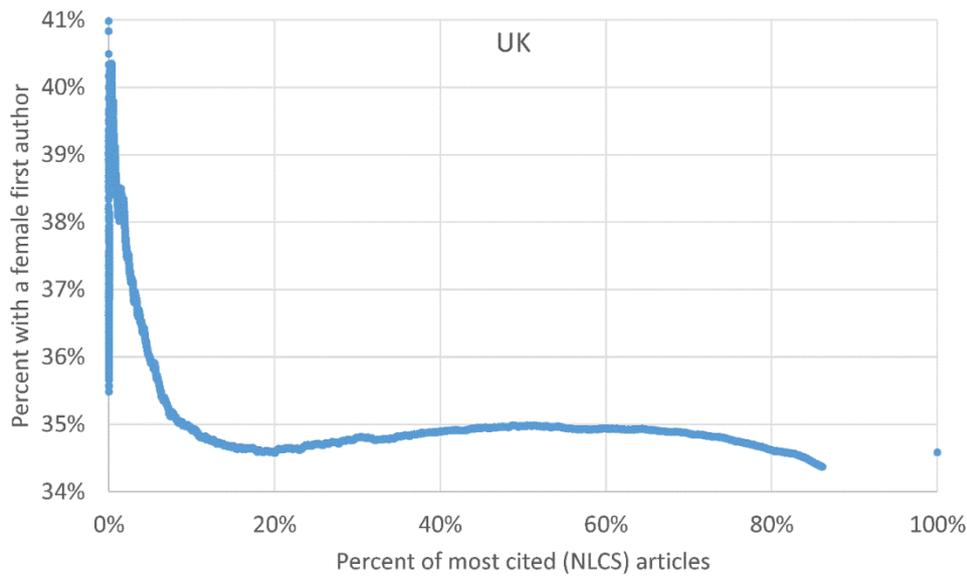

Figure 13. The cumulative percentage of female first authored **UK** Scopus-indexed journal articles 1996-2018 in each top cited percentile (using NLCS values).

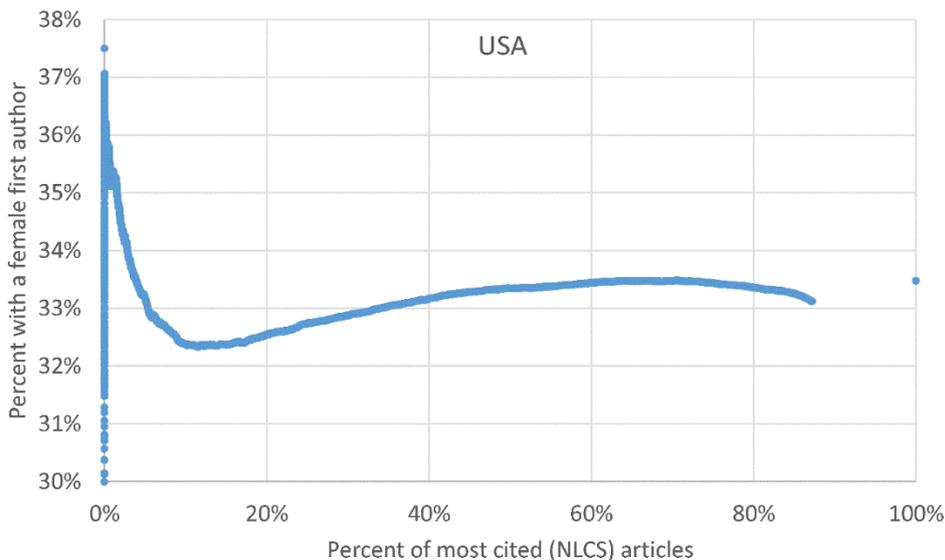

Figure 14. The cumulative percentage of female first authored **US** Scopus-indexed journal articles 1996-2018 in each top cited percentile (using NLCS values).

## Discussion

The results are limited by the Scopus journal classification scheme. An Elsevier decision to add or remove a large journal from a category with an unusual gender distribution for that category might impact the overall MNLCS score. Previous research comparing different classification schemes suggests that this should not have a large effect on the results (Thelwall, 2018a). This is nevertheless an important hidden factor, and so small differences should be interpreted cautiously. The data for the most recent three years could be misleading because of the possibility that they are unduly influenced by self-citations, as well as the lower MNLCS accuracy due to the higher proportion of uncited articles and outliers (high kurtosis values). Self-citation rates vary by country (Deschacht & Maes, 2017) and so may affect the seven nations differently. The results before 2014 should not be affected by outliers, however, except for a small number of years, because the kurtosis values are moderate in almost all cases. The results are also incomplete due to the third of authors with an undetected gender. These could affect the results, especially if they are from minority cultures with different gender roles or expectations. Ignoring the influence of authors after the first is also a limitation, especially in fields where the last author is senior and influential in choosing the research topic (Duffy, 2017; Marschke, Nunez, Weinberg, & Yu, 2018). The results may be influenced by the genders of contributions of authors after the first. Citations reflect only one type of research impact (Priem, 2014) and are affected by many apparently irrelevant considerations (Didegah & Thelwall, 2013) and process factors (Klitzing, Hoekstra, & Strijbos, 2019). Finally, the conclusions should not be extrapolated to other countries because gender roles vary widely on an international scale, including within academia (e.g., Othman & Latih, 2006).

To check whether the results would be different for research with a single author (and therefore unambiguously the work of a single gender), the analyses were repeated for single author journal articles. The results are available in the form of versions of Figures 1 to 7 and Table 2 in the supplementary material file. In all countries there was a slightly smaller proportion of females for solo-authored research than for first authoring all articles. In all countries, the average citation impact of solo-authored articles was lower than for all articles, but the same gender difference patterns remained: female solo research had a higher average citation impact than male solo research in six out of seven countries, with the same exception, the USA. Thus, the main conclusions also hold for solo research.

A previous paper comparing male and female MNLCS in the USA for articles published in 2015 found a tiny female citation advantage (Thelwall, 2018a) whereas Figure 7 shows a tiny male citation advantage. The difference is due to the extended list of gendered first names used in the current paper: if Figure 7 for the USA is recreated with the previous gendered name list then (with the updated Scopus citation counts) there is a tiny female citation advantage. The results contrast with the larger male global citation advantages previously found (Larivière et al., 2013), which seems to be due to the different field normalisation process used (Thelwall, 2018a).

The results contrast previous characterisations of the most highly cited papers as mainly male-authored (Baltussen & Kindler, 2004; Graham, Pratt, Lee, & Cullen, 2019; Schisterman, Swanson, Lu, & Mumford, 2017; Wong, Tan, & Sabanayagam, 2019). This may be due to field differences or the relatively fine-grained subject classification used here so that fast publishing specialisms are less likely to have an advantage, as well as advantages for older articles that have had more time to be cited. It is also possible that a focus on a small fraction of a percentile of articles for a top list would be male-biased even if the top few percentiles include more female first-authored papers than average (Figures 8, 9, 10, 14). The field normalisation process used, in conjunction with combining results from multiple fields may also affect the results because it alters the balance between the average and individual values in a non-linear way.

What can explain the higher average female citation impact results for most countries? The results are more impressive given the higher self-citation rates of males (King et al., 2017). A bias against citing males seems unlikely given historical sexism against women in academia. There is some evidence of gender homophily in citing (Potthoff & Zimmermann, 2017) but this would favour males, who are a majority in Scopus. There is little evidence for gender differences in innate intellectual capabilities (Hines, 2011), so biological superiority for women is not currently a credible explanation for the core tasks involved in research (see also: Ceci & Williams, 2011). As mentioned above, there is strong evidence from vocational psychology (although mainly from the USA, where there are not substantial gender differences in the overall results above) that female career choices are more likely to align with societal or communal goals (Diekman, et al., 2017). This is evident in academia through gender differences in academic field specialisms (e.g., more females in nursing, more males in maths: Thelwall et al., 2019). Thus, it is plausible that female-first authored research is more impactful (i.e., not just higher citation impact), at least in most large English-speaking nations, because females are socialised to want to carry out work that is meaningful to society. This may occur by choosing more useful research topics *within* a field, for example, or by the lead author conducting more activities to translate their findings into practical value. This hypothesis has no direct evidence to support it yet, other than from one Mendeley study for education (Thelwall, 2018a).

## Conclusions

The results show no evidence of a historical citation disadvantage against female first-authored research in large English-speaking nations, even in the late 1990s where the proportion of women in academia was substantially lower than today. Thus, previously hypothesised historical citation disadvantages or bias cannot explain continuing gender imbalances in academia half a century after employment sex discrimination was outlawed, and other explanations must be addressed instead. These include historical and current gender disadvantages or bias in other contexts, and systemic biases, including gender role expectations and the greater share of unpaid caring activity for females, as well as different life choices (Ceci & Williams, 2011).

In contrast, the higher average citation impact of female first-authored research in most of the large English-speaking countries examined points to the possibility that each female first-authored output is more valuable. This is supported by some evidence that female first-authored research may have more non-academic impacts (Thelwall, 2018a). This and the greater female ability to write high impact research (e.g., in the top 3.5% for impact for the USA) *may* be a side-effect of a tendency for female socialisation processes that lead to valuing societal or communal goals within careers. Appointment, tenure, funding and promotion committees should therefore evaluate the likely impact of all candidates' research if they want to avoid unintentional bias against female candidates. For example, if two candidates were otherwise similar then a committee should prefer the candidate with research that they judged to be more likely to have an impact on science or society. The results here show that this would tend to favour females in at least six of the seven English speaking countries examined and may be gender neutral in the USA, but positive if higher impact outputs were focused on. This is an additional factor that should be considered alongside the known systemic issues, such as the importance of allowing for career breaks and creating a supportive environment for those with additional personal life responsibilities.